\documentclass[letterpaper]{aa520}

\usepackage{txfonts}
\usepackage{graphicx}
\usepackage{natbib}
\usepackage{amssymb}

\bibpunct{(}{)}{;}{a}{}{,} 

\begin{document}

\title{Intermediate scale structure of the interstellar medium towards 
       NGC~6231 in Sco~OB1 with FUSE\thanks{Based on observations 
       made with the NASA-CNES-CSA Far Ultraviolet Spectroscopic Explorer. 
       FUSE is operated for NASA by the Johns Hopkins University under 
       NASA contract NAS5-32985.}
}
\author{O. Marggraf
        \and
        H. Bluhm
        \and
        K.S. de Boer
} 

\offprints{O. Marggraf,\\
           \email{marggraf@astro.uni-bonn.de}}

\institute{
Sternwarte der Universit\"at Bonn, Auf dem H\"ugel 71, D-53121 Bonn, Germany
}

\date{Received 16 June 2003 / Accepted 2 December 2003}

\titlerunning{Intermediate scale structure of the ISM towards NGC~6231}
\authorrunning{O. Marggraf et al.}

\abstract{
The FUSE far-ultraviolet interstellar spectra toward seven targets in 
\object{NGC\,6231} show that the molecules H$_2$, HD, and CO as well as various
atomic species are distributed in several clouds.
The main absorption component found on the sight lines lies in the Lupus cloud
region at a distance of about 150~pc, and there is a weaker second one,
presumably in the \object{Sco OB1} shell surrounding \object{NGC\,6231} 
($d\simeq1.8$~kpc).
H$_2$ excitation modelling is used to constrain the radiation field and the
density in the gas;
HD is used to estimate the abundance of H$^+$.
The small angular separation of the target stars allows column density 
variations to be probed over the field of view, on scales of $0.05$~pc in the
case of the Lupus cloud distance.
They are $40\%$ for H$_2$ and $60\%$ for \ion{H}{i}.
A rather strong radiation field inside the molecular clouds suggests a 
separation of the gas into smaller cloudlets also along the line of sight.
}

\maketitle


\section{Introduction}

When observing the interstellar medium (ISM) using absorption line 
spectroscopy, one has to live with the restriction that the data gives just 
pencil beam information along single lines of sight towards selected background
sources. 
Absorption line analysis will never provide as wealthy structural 
information as the analysis of emission lines or imaging.

However, one can get closer to this goal by choosing groups of close together 
background sources, in this way enlarging the information density per angular 
element.

Open star clusters, rich in young UV bright background sources, play an
important role in such projects when dealing with far-UV (FUV) absorption line 
spectroscopy of the ISM. 
In particular, the open cluster \object{NGC\,6231} in the 
\object{Sco OB1} association provides at least seven early-type stars in its 
central region which are suitable as background sources for FUV spectroscopy, 
and it still shows moderate extinction in this wavelength region for a far-away
object in the Milky Way disk.
The cluster is located in the Galactic plane ($l=343.46\degr$, $b=+1.17\degr$),
at a distance of $\sim1.8$~kpc \citep{raboud97}.
The $E(B-V)$ values for its stars vary between $0.43$ and $0.46$~mag 
\citep[][see Table~\ref{tab_targets}]{baume99}.

We will analyse the sight lines towards these seven background sources
with respect to absorption by the Galactic interstellar medium, paying
particular attention to the variation in the physical parameters of the gas 
over small angular scales.
The background sources span a total area of $5.8\arcmin \times 3.1\arcmin$.
At the distance of the cluster this means a linear extent of the gas 
of about $3.0$~pc\,$ \times 1.6$~pc.
Part of the absorbing gas will very probably be located more nearby, resulting
in an even higher spatial resolution.

The spectral range of the FUSE instrument covers the Lyman and Werner bands 
of molecular hydrogen H$_2$ and of deuterated hydrogen HD, the 
\ion{H}{i} Lyman series from Ly\,$\beta$ on bluewards, the CO E-X band, 
and various metals.
We are thus able to derive physical parameters for the gas like excitation
temperatures and densities from excitation and column density ratios.
Sect.~\ref{sect_obs} introduces the data themselves and the analysis methods
used;
Sect.~\ref{sect_res} gives an overview of the results and a 
discussion of their reliability.
Sect.~\ref{sect_interp} contains the interpretation, followed by
concluding remarks in Sect.~\ref{sect_conc}.


\section{Instrumentation, observation, and data reduction}

\label{sect_obs}

\begin{table*}[ht!]
\centering
\caption{List of the targets and observational parameters. 
         The spectrum of \object{HD\,152270} was observed in $5$ separate 
         pointings, thus no unique ID is given.
         The $E(B-V)$ values are from \citet{baume99}.
        }
\label{tab_targets}
\begin{tabular}{llcclccc}
\hline\hline\noalign{\smallskip}
Target       & Sp.Type & RA (2000) & Dec (2000) & FUSE ID     & Exp. date & Exp. time [s] & $E(B-V)$ \\
\noalign{\smallskip}\hline\noalign{\smallskip}
\object{HD 152200}    & O9.5III & 16 53 51.69 & -41 50 33.1 & B0250101000 & 2001-08-09 17:48:00 & 5367 & 0.43 \\
\object{HD 152219}    & O9.5III & 16 53 55.65 & -41 52 51.8 & B0250201000 & 2001-08-10 13:46:00 & 4905 & 0.46 \\
\object{HD 152234}    & B0.5Ia  & 16 54 01.88 & -41 48 23.4 & B0250301000 & 2001-08-11 18:22:00 & 5123 & 0.45 \\
\object{HD 152249}    & O9.5Iab & 16 54 11.66 & -41 50 57.8 & B0250401000 & 2001-08-12 02:26:00 & 4947 & 0.44 \\
\object{HD 326329}    & O9V     & 16 54 14.06 & -41 50 09.0 & B0250501000 & 2001-08-09 12:45:00 & 5485 & 0.44 \\
\object{CPD -47 7712} & B0IV    & 16 54 00.44 & -41 52 44.0 & B0250601000 & 2001-08-09 04:24:00 & 9611 & 0.46 \\
\object{HD 152270}    & WC7     & 16 54 19.70 & -41 49 11.5 & B013020x000 & 2001-07-08 23:00:48 & 2950 & 0.46 \\
\noalign{\smallskip}\hline
\end{tabular}
\end{table*}

\subsection{Data observation and preparation}

The FUSE instrument consists of four coaligned telescopes with Rowland
spectrographs and two microchannel plate detectors.
The spectral range of operation is between 905 and 1187~\AA, at a resolution 
of about $\lambda/\Delta\lambda\simeq15000$ or 20~km\,s$^{-1}$ (FWHM).
An extended overview of the instrumentation and the on-orbit performance of the
satellite is found in \citet{moos00} and in \citet{sahnow00}.

\begin{figure}[ht!]
\resizebox{\hsize}{!}{\includegraphics{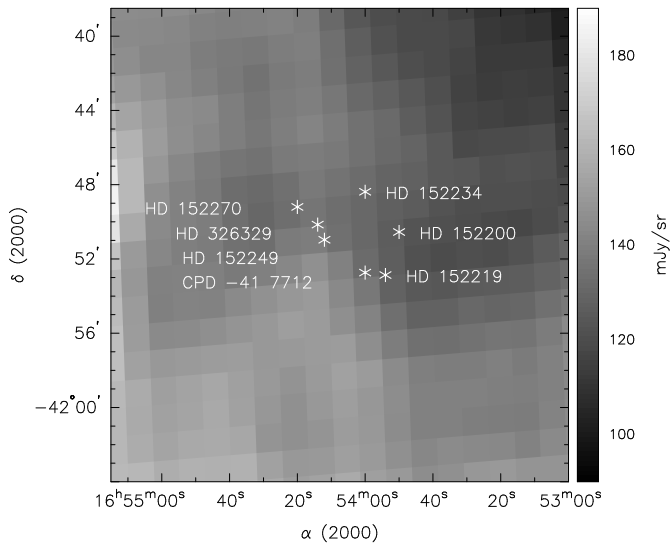}}
\caption[]{Positions of the target stars. The background shows the 
           $60$~$\mu\rm{m}$ IRAS emission map (see intensity scale at right).
          }
\label{fig_positions}
\end{figure}

The open cluster \object{NGC\,6231} was selected as a target with the
criteria in mind of 
{\sl a)} a cluster with a number of hot early type stars 
(earlier than about B3, see Table~\ref{tab_targets}) providing clean UV 
continua without perturbing stellar atmospheric absorption and
{\sl b)} a moderate colour excess so that the UV flux is not diminished
by a too strong interstellar grain absorption.
Table~\ref{tab_targets} lists the FUSE targets and the observational
parameters. 
As an additional benefit the individual background sources are all separated
by at least $1\arcmin$,
thus we are able to use the large LWRS aperture ($30\arcsec \times 30\arcsec$)
for observation. 
The separation is large enough to assure that always only one UV bright star
is located within the aperture.
A map of the \object{NGC\,6231} region and the locations of the FUSE targets
is given in Fig.~\ref{fig_positions}.

We use the CalFUSE standard reduction package (version 2.0.5) for the basic
data reduction and calibration.
The different exposures for each target are cross correlated and coadded for
each detector channel.
The coadded spectra are then rebinned to a binsize of $4$, which results in a 
resolution of 0.028~\AA\ per bin (or $8.4$~km\,s$^{-1}$ at 1000~\AA), 
still well below the instrument's spectral resolution.

\subsection{Fitting procedure}

\begin{figure*}[t!]
\includegraphics[width=18cm]{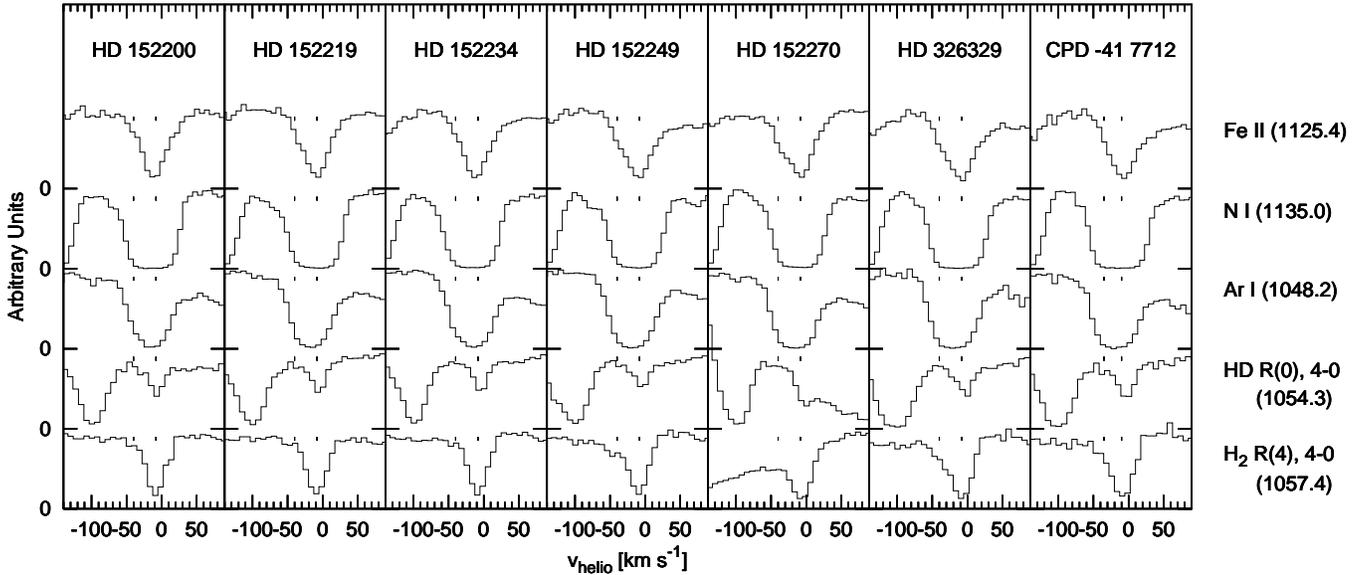}
\caption[]{Absorption lines from the seven target star spectra.
           The sample portions are binned (to $\sim8.4$~km\,s$^{-1}$), 
           but not normalised to a continuum.
           The \ion{Ar}{i} line at 1048~\AA\, is located in the left wing of
           a strong H$_2$ absorption feature.
           Stellar wind features affect the continuum in the WR star 
           \object{HD\,152270}.
           For some sight lines (\object{HD\,152249}, \object{HD\,152270}, 
           \object{HD\,326329}) a second blueward velocity component is
           clearly visible, while it is much weaker in the others.
           The dashes above the spectra mark the positions of the main 
           velocity component at $\sim-8$~km\,s$^{-1}$ and of the weaker
           second component at $-40$~km\,s$^{-1}$.
           At the left end of the box of the \ion{N}{i} line the transition of
           \ion{N}{i} (1134.4~\AA) can be seen, shortwards of the HD\,R(0) line
           a blend of the H$_2$ transitions R(3), 4-0 and R(6), 5-0. 
          }
\label{7plot}
\end{figure*}

The analysis of the absorption lines is only possible if one applies full Voigt
profile fitting to the absorption as a whole.
Strong damping wings and heavy blends especially of the molecular lines 
hamper the identification of single absorption lines and the determination 
of equivalent widths for them;
blends of lines of different species or excitation levels are common.

The Doppler parameter of the absorption $b$ was estimated beforehand
using a standard curve of growth analysis, but on a restricted set of 
absorption lines.
This was done separately for the molecular and for the atomic gas since they 
possibly exist in different environments.
For the molecules we used the H$_2$ transitions of the J3, J4, and J5 
excitation states which, for our sight lines, define the full flat part of the
curve of growth (as preparatory plots showed).
These lines thus tend to have only moderate damping wings, if at all, and
readily provide equivalent widths.
For the metals we used the full series of available iron absorption lines 
except the few that are obviously blended.

\begin{figure*}[t!]
\centering
\includegraphics[width=17cm]{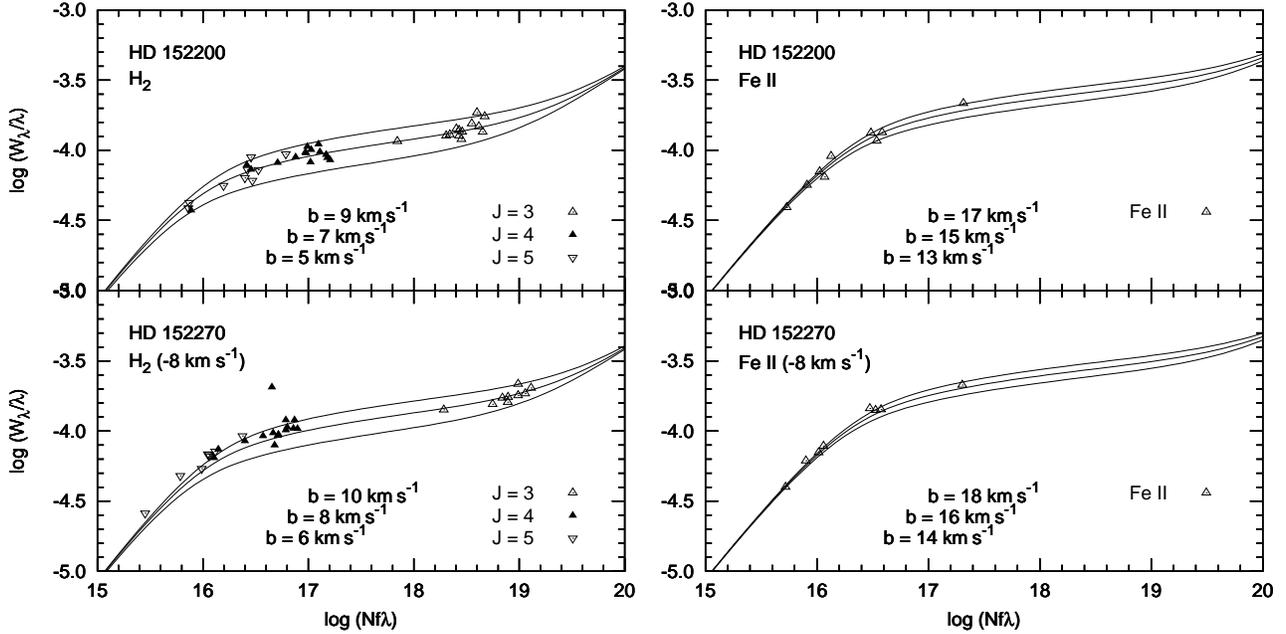}
\caption[]{Sample curves of growth.
           Curves of growth were only used to determine an initial preliminary
           $b$ value for the fitting process; 
           this $b$ value was then subject to some adjustment during the fit.
           Shown are single-component curves of growth for the $b$ values that
           resulted from the fit, but the equivalent widths plotted
           are sums over the full absorption features, i.e. sums over the 
           equivalents widths of the individual blended velocity components. 
           For this reason a discrepancy between equivalent widths and
           the curve of growth is visible for the curve of the two-component
           sight line \object{HD\,152270}.
          }
\label{fig_CoGs}
\end{figure*}

The $b$ values so estimated were then taken as input parameters for an
interactive $\chi^2$ fit to the FUSE spectra.
We generate a full theoretical Voigt profile absorption line spectrum over the
FUSE wavelength range, fitting the radial velocity of the absorbing gas and the
column densities of all major absorbing species (including \ion{H}{i}) at once.
A continuum level is estimated by eye.
This way we can also handle the extended damping wings and blends in the lines
in particular of the lower H$_2$ excitation levels and also the various 
inter-species blends.
In the process of fitting the initially estimated $b$ values were adjusted 
as well, if nessessary.
Fig.~\ref{fig_CoGs} shows a sample of curves of growth using the final $b$ 
values and column densities.

The lines' transition parameters are taken from \citet{abgrall93a,abgrall93b}
for the H$_2$ Lyman and Werner band (utilising the electronic version of the
data provided by 
S.\,R.\,McCandliss\footnote{WWW URL: http://www.pha.jhu.edu/$^\sim$stephan/h2ools2.html}), 
\citet{morton75,morton78} and \citet{wright79} for HD, and \citet{morton94} and
\citet{federman01a} for CO.
The metal transitions are from \citet{morton03}, with some values for Fe taken
from \citet{howk00}.
The full set of transitions is included in the fitting process, however the 
weakest lines do not neccessarily contribute to the result. 
A small number of stronger transitions had to be disregarded due to 
unidentified blends or detector effects.

We use the wavelength/velocity calibration that is provided by the 
FUSE 1A\,LiF spectra.
Although the FUSE wavelength calibration is, at present, known to be 
problematic \citep[see, e.g., the discussion in Par.~3.4 of ][]{wakker03}
comparison with the velocity structure observed in high resolution 
optical spectra \citep{crawford95, crawford01} shows a high similarity.
We thus assume that the wavelength scale defined in the 1A\,LiF (and also the
2A\,LiF) spectra is very close to reality.
Corrections would have had to be applied to the spectra of some of the other
detectors, though, but they are not used in this work.

\begin{figure*}[t!]
\sidecaption
  \includegraphics[height=12cm,angle=90]{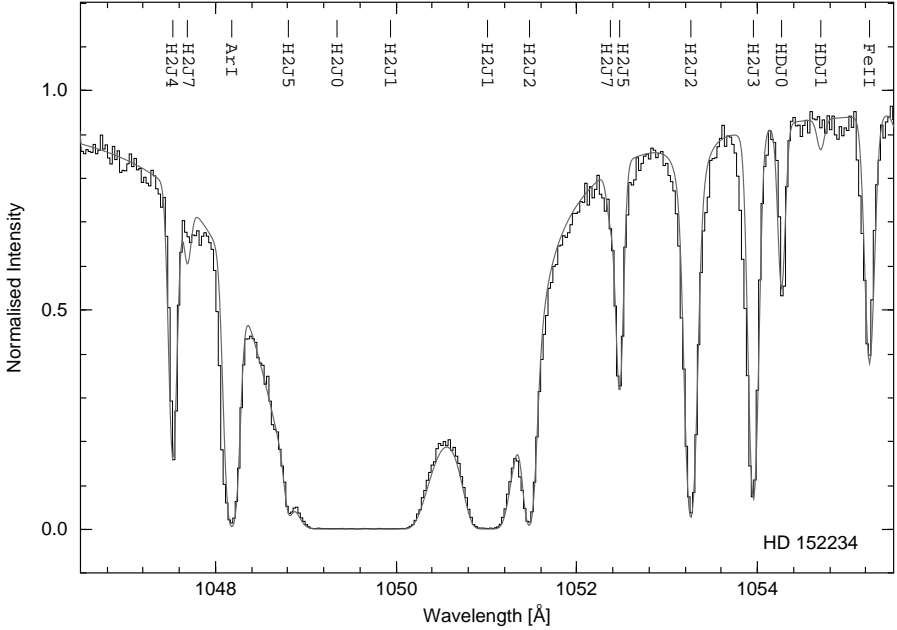}
  \caption{~Sample portion of the \object{HD\,152234} spectrum with a
           one-component fit, in which the positions of the fitted transitions
           are indicated.
           The continuum count rate in this region of the spectrum is ca. 
           $6000$ (binned).
           The full wealth of H$_2$ transitions mainly of the (4-0) 
           vibrational states is visible.
           Also visible is a strong HD line of the $J=0$ state 
           (HD R(0), 4-0).
           For the $J=1$ state of HD we measure only upper limits.
           The \ion{Ar}{i} transition at $1048$~\AA\, (see also 
           Fig.~\ref{7plot}) is affected by some unknown absorption on its 
           left wing, maybe by a weak second component bluewards. 
           Also other lines hint at the presence of a weak second component.
          }
  \label{fig_spectrum}
\end{figure*}

Fig.~\ref{7plot} shows samples of various unblended absorption lines on the 
seven sight lines.
In Fig.~\ref{fig_spectrum} a sample of typical heavily blended H$_2$ 
absorption structures is given, together with the full Voigt profile fit.

\subsection{Error estimates}

Errors for the column densities are determined as $3\,\sigma$ deviations from
the fit.
In addition to these, we have to take care of additional uncertainties from
four sources which are not included in the fitting process itself.

\vspace{1mm}\noindent
{\it The FWHM:}

The instrumental FWHM in our spectra seems to change over the FUSE detector 
segments, with the 1A\,LiF spectra having a slightly higher FWHM 
(24~km\,s$^{-1}$, in the case of \object{HD152200}) than the 2A\,LiF spectra 
(22~km\,s$^{-1}$). 
This adds some uncertainty to the column density fit, however it measurably
affects only the fits for \ion{N}{i} (near $\lambda=1134$~\AA) and for the 
$J=3$ to $5$ levels of H$_2$.
In general, the variation of column density with FWHM is in the range of the
small error from the $\chi^2$ fit.
For the few elements mentioned one may get up to three times this value.

The three H$_2$ levels are the ones which we use in fixing the $b$ value
because of their location on the flat part of the curve of growth.
The curve of growth procedure, however, deals with equivalent widths only and
is thus independent from changes in FWHM.
Thus, the origin of this particular variation on changing FWHM is rather caused
by the general uncertainty in fitting elements on the flat part of the curve 
of growth, not by the uncertainty in FWHM itself.

The lines of \ion{N}{i} are in general difficult to fit due to their close
triplet nature, they are highly sensitive to changes in any fitting parameter,
including FWHM.

\vspace{1mm}\noindent
{\it The continuum:}

Another question is the dependency of the resulting column densities 
on the assumed continuum. 
There seems to be no proper way to quantify this dependency. 
Of course, elements with only few transitions are more sensitive to 
changes in the local continuum of a particular line.
Among those, the saturated lines of especially \ion{O}{i} and \ion{N}{i}
are more affected, since the fit is here mainly sensitive to changes in the 
flanks of the absorption profile, while the profiles themselves are not very
sensitive to changes in column density.

Overall, we find a variation in column density of about the size of the 
$\chi^2$ fit errors (about twice that value for \ion{O}{i} and \ion{N}{i}) 
if one varies the local continuum by about $10\%$ of the originally assumed 
level.
The resulting continua are outside the range of what would be considered a 
valid continuum for our spectra; therefore this gives a good estimate of the
uncertainty introduced by the continuum selection.

\vspace{1mm}\noindent
{\it The b value:}

The $b$ value is the largest and also the easiest to underestimate 
contributor of errors, whether one uses the conventional curve of growth 
technique or any sophisticated fitting procedure.
Even in high-quality spectra it can be difficult to fix. 

For the metals, only iron is suitable for determining $b$, all other elements 
having too few usable transitions in the FUSE spectral range.
As a consequence, some of the latter are also more error-prone.
In our case, the column densities in particular of oxygen and nitrogen are 
extremely sensitive to changes in $b$, with their line strengths at the level
of the flat part of the curve of growth.
The less abundant elements are located on the linear part of the curve and
are thus well defined, 
but for the same reason they cannot be used to fix the $b$ value.

The various H$_2$ excitation states are located all over the curve of growth,
but only the slightly saturated middle levels define the $b$ value for
the molecular gas.
As for the metals, their column densities again are most sensitive to changes
in the $b$ value, while the equivalent widths of the transitions and their 
profiles change only marginally.
The quality of the fit therefore does not suffer much and does not provide any
measure for the quality of the $b$ value.

Fitting the spectrum with different $b$ values gives an impression of the
variation. 
We varied the originally determined $b$ value by $\pm1$~km\,s$^{-1}$.
Column densities for elements having only absorption lines with a strength of
the flat part of the curve of growth changed by up to $0.5$~dex, for elements 
low on the linear part or high in the damping part about $0.1$~dex or even
less.
As error from the $b$ value uncertainty we assume, for each element, the mean
deviation for $\chi^2$ fits with $b\pm1$~km\,s$^{-1}$.

\vspace{1mm}\noindent
{\it The stellar contribution:}

Stellar atmospheric features are relatively few in number and then easily 
visible, mostly wide features. 
Strong wind profiles are found, especially in the Wolf-Rayet star 
\object{HD\,152270}.
We tested modelling the stellar spectra using Kurucz atmospheric models, 
without a satisfying result.
Broad stellar features are therefore included by eye in the assumed continuum.
If interstellar lines are affected by narrow stellar atmospheric lines in a way
that such a continuum cannot unambiguously be defined, 
these regions are excluded from the fit.

\vspace{1mm}

The individual errors from all these sources -- $\chi^2$ fit and uncertainties
in FWHM, continuum, and $b$ value -- add up to the given total errors.
In most cases, the errors from the fit itself are very small, so the larger
contribution comes from the uncertainty in the continuum, in the FWHM, and
especially the $b$ value.

\begin{table*}[ht!]
\centering
\caption{List of column densities $\log N$~[$cm^{-2}$] towards NGC~6231.
         FWHM, radial velocities $v$, and Doppler widths $b$ are in 
         km\,s$^{-1}$.
         Upper limits are at least $3\sigma$. 
         The $b$ values are ``effective'' values for the measured absorption 
         lines and should be lower for individual unresolved subcomponents.
        }
\label{tab_CDs}
\setlength{\tabcolsep}{1.70mm}
\begin{tabular}{lrrrrrrrrrr}
\hline\hline\noalign{\smallskip}
           & 152200 & 152219 & 152234 & 152249 & & 152270 & & 326329 & & -41\,7712 \\
\noalign{\smallskip}\hline\noalign{\smallskip}
FWHM & 24 & 24 & 24 & 22 & 22 & 22 & 22 & 22 & 22 & 22 \\
\noalign{\smallskip}\hline\noalign{\smallskip}
$v$ & $-8$ & $-8$ & $-8$ & $-8$ & $-40$ & $-8$ & $-40$ & $-8$ & $-35$ & $-10$ \\
\noalign{\smallskip}\hline\noalign{\smallskip}
$b$ & 7.0 & 7.0 & 7.0 & 6.0 & 6.0 & 8.0 & - & 7.0 & 7.0 & 7.0 \\
\noalign{\smallskip}\hline\noalign{\smallskip}
H$_2$, $J=0$ & $20.14^{+0.02}_{-0.02}$ & $20.10^{+0.02}_{-0.02}$ & $20.21^{+0.02}_{-0.03}$ & $20.00^{+0.02}_{-0.05}$ & -                       & $20.22^{+0.02}_{-0.04}$ & - & $20.08^{+0.02}_{-0.02}$ & -                        & $20.25^{+0.02}_{-0.02}$ \\
H$_2$, $J=1$ & $19.97^{+0.02}_{-0.02}$ & $20.00^{+0.02}_{-0.02}$ & $20.01^{+0.02}_{-0.02}$ & $19.89^{+0.02}_{-0.02}$ & $>16.00$                & $20.03^{+0.02}_{-0.02}$ & - & $19.95^{+0.02}_{-0.02}$ & $<15.40^{+0.49}_{-0.49}$ & $20.11^{+0.02}_{-0.02}$ \\
H$_2$, $J=2$ & $18.12^{+0.27}_{-0.28}$ & $18.39^{+0.11}_{-0.11}$ & $18.25^{+0.16}_{-0.16}$ & $18.42^{+0.08}_{-0.08}$ & $14.70^{+0.32}_{-0.29}$ & $18.68^{+0.05}_{-0.05}$ & - & $18.58^{+0.09}_{-0.07}$ & $<14.60^{+0.32}_{-0.32}$ & $18.72^{+0.06}_{-0.06}$ \\
H$_2$, $J=3$ & $17.40^{+0.45}_{-0.45}$ & $17.56^{+0.43}_{-0.43}$ & $17.18^{+0.60}_{-0.21}$ & $17.62^{+0.39}_{-0.39}$ & $14.90^{+0.26}_{-0.36}$ & $17.84^{+0.39}_{-0.39}$ & - & $17.97^{+0.36}_{-0.36}$ & $15.12^{+0.06}_{-0.06}$ & $18.26^{+0.18}_{-0.18}$ \\
H$_2$, $J=4$ & $15.96^{+0.16}_{-0.16}$ & $15.91^{+0.18}_{-0.18}$ & $15.94^{+0.16}_{-0.16}$ & $15.88^{+0.14}_{-0.14}$ & $14.30^{+0.17}_{-0.23}$ & $15.65^{+0.12}_{-0.12}$ & - & $<15.96^{+0.30}_{-0.30}$ & $<14.48^{+0.25}_{-0.25}$ & $16.29^{+0.14}_{-0.16}$ \\
H$_2$, $J=5$ & $15.40^{+0.10}_{-0.10}$ & $15.14^{+0.07}_{-0.07}$ & $15.22^{+0.08}_{-0.08}$ & $14.95^{+0.04}_{-0.04}$ & $14.20^{+0.13}_{-0.13}$ & $14.99^{+0.03}_{-0.04}$ & - & $15.18^{+0.07}_{-0.08}$ & $<14.46^{+0.48}_{-0.48}$ & $15.22^{+0.09}_{-0.09}$ \\
H$_2$, $J=6$ & $<14.15$                & $<14.00$                & $14.09^{+0.05}_{-0.05}$ & $<14.00$                & -                       & $<14.15$                & - & $<14.15$                & -                       & $<14.20$                 \\
H$_2$, $J=7$ & $14.08^{+0.04}_{-0.04}$ & $14.06^{+0.02}_{-0.02}$ & $14.12^{+0.02}_{-0.02}$ & $<14.20$                & -                       & $14.02^{+0.04}_{-0.02}$ & - & $<14.10$                & -                        & $<14.20$                \\
\noalign{\smallskip}\hline\noalign{\smallskip}
HD, $J=0$ & $14.64^{+0.09}_{-0.10}$ & $14.59^{+0.06}_{-0.05}$ & $14.62^{+0.05}_{-0.04}$ & $14.51^{+0.06}_{-0.06}$ & -                       & $14.55^{+0.07}_{-0.05}$ & - & $14.61^{+0.10}_{-0.09}$ & -                        & $14.71^{+0.09}_{-0.10}$ \\
HD, $J=1$ & $<14.00$                & $<13.90$                & $<13.95$                & $<13.90$                & -                       & $<14.15$                & - & $<14.20$                & -                        & $<14.30$ \\
\noalign{\smallskip}\hline\noalign{\smallskip}
CO$^1$ & $13.85^{+0.06}_{-0.12}$ & $13.78^{+0.08}_{-0.04}$ & $13.91^{+0.03}_{-0.02}$ & $13.82^{+0.05}_{-0.01}$ & -                       & $13.95^{+0.05}_{-0.02}$ & -                       & $13.99^{+0.11}_{-0.05}$ & -                      & $13.89^{+0.11}_{-0.06}$ \\
CO$^2$ & $>13.70$ & $>13.63$ & $>13.69$ & $>13.67$ & - & $>13.60$ & - & $>13.79$ & - & $>13.87$ \\
\noalign{\smallskip}\hline\hline\noalign{\smallskip}
$v$  & $-6$ & $-2$ & $-6$ & $-1$ & $-40$ & $+2$ & $-40$ & $-3$ & $-40$ & $-4$ \\
\noalign{\smallskip}\hline\noalign{\smallskip}
$b$  & 9.0 & 9.0 & 10.0 & 10.0 & - & 9.0 & - & 10.0 & - & 9.0  \\
\noalign{\smallskip}\hline\noalign{\smallskip}
\ion{C}{i}\,$(^3P_0)$ & $15.15^{+0.05}_{-0.05}$ & $15.05^{+0.03}_{-0.06}$ & $15.17^{+0.02}_{-0.02}$ & $15.05^{+0.07}_{-0.07}$ & - & $15.00^{+0.15}_{-0.15}$ & - & $15.14^{+0.05}_{-0.04}$ & - & $<15.30^{+0.05}_{-0.05}$ \\
\ion{C}{i}\,$(^3P_1)$ & $14.86^{+0.03}_{-0.03}$ & $<14.90$                & $15.00^{+0.03}_{-0.03}$ & $<14.75$                & - & $<14.80$                & - & $14.97^{+0.06}_{-0.04}$ & - & $<15.00$                \\
\ion{C}{i}\,$(^3P_2)$ & $<14.60$                & $<14.60$                & $<14.75$                & $<14.45$                & - & $<14.60$                & - & $<14.55$                & - & $<14.60$                \\
\noalign{\smallskip}\hline\hline\noalign{\smallskip}
$v$  & $-14$ & $-11$ & $-13$ & $-8$ & $-40$ & $-8$ & $-40$ & $-8$ & $-35$ & $-10$ \\
\noalign{\smallskip}\hline\noalign{\smallskip}
$b$  & 15.0 & 16.0 & 14.0 & 14.0 & 5.0 & 16.0 & 5.0 & 12.0 & 7.0 & 16.0 \\
\noalign{\smallskip}\hline\noalign{\smallskip}
\ion{H}{i}   & $21.23^{+0.02}_{-0.02}$ & $21.38^{+0.02}_{-0.02}$ & $21.32^{+0.02}_{-0.02}$ & $21.40^{+0.08}_{-0.02}$ & -                       & $21.44^{+0.10}_{-0.02}$ & - & $21.36^{+0.02}_{-0.02}$ & -                        & $21.37^{+0.02}_{-0.02}$ \\
\noalign{\smallskip}\hline\noalign{\smallskip}
\ion{O}{i}   & $18.52^{+0.42}_{-0.49}$ & $18.35^{+0.37}_{-0.43}$ & $18.95^{+0.44}_{-0.45}$ & $17.10^{+0.30}_{-0.37}$ & -                       & $17.05^{+0.25}_{-0.34}$ & $18.30^{+0.84}_{-0.87}$ & $<18.95$                & $19.00^{+1.05}_{-0.43}$ & $18.41^{+0.42}_{-0.58}$ \\
\ion{Si}{ii} & $<16.85$                & $17.63^{+0.36}_{-0.48}$ & -                       & -                       & -                       & $17.35^{+0.48}_{-1.26}$ & $<16.95$                & -                       & -                       & $17.49^{+0.51}_{-1.59}$ \\
\ion{Fe}{ii} & $15.23^{+0.03}_{-0.03}$ & $15.18^{+0.04}_{-0.03}$ & $15.23^{+0.03}_{-0.04}$ & $15.16^{+0.02}_{-0.02}$ & $14.41^{+0.07}_{-0.07}$ & $15.15^{+0.02}_{-0.02}$ & $14.37^{+0.08}_{-0.08}$ & $15.22^{+0.04}_{-0.04}$ & $14.53^{+0.04}_{-0.04}$ & $15.18^{+0.03}_{-0.03}$ \\
\ion{Ar}{i}  & $15.05^{+0.13}_{-0.13}$ & $15.43^{+0.17}_{-0.18}$ & $15.13^{+0.15}_{-0.15}$ & $14.69^{+0.12}_{-0.06}$ & $>15.20$?               & $14.67^{+0.09}_{-0.12}$ & $14.50^{+0.83}_{-0.53}$ & $14.89^{+0.23}_{-0.37}$ & $<16.55$                & $15.56^{+0.20}_{-0.22}$ \\
\ion{N}{i}   & $17.40^{+0.37}_{-0.42}$ & $17.20^{+0.31}_{-0.32}$ & $17.45^{+0.47}_{-0.43}$ & $17.00^{+0.38}_{-0.34}$ & $>16.60$?               & $17.30^{+0.49}_{-0.49}$ & $16.50^{+0.69}_{-0.69}$ & $17.30^{+0.50}_{-0.37}$ & $<17.45$                & $17.35^{+0.36}_{-0.44}$ \\
\ion{P}{ii}  & $14.24^{+0.07}_{-0.08}$ & $14.25^{+0.07}_{-0.07}$ & $14.18^{+0.05}_{-0.05}$ & $14.10^{+0.04}_{-0.04}$ & $13.76^{+0.23}_{-0.24}$ & $14.16^{+0.04}_{-0.06}$ & $14.14^{+0.04}_{-0.06}$ & $14.17^{+0.15}_{-0.08}$ & $13.85^{+0.25}_{-0.25}$ & $14.25^{+0.09}_{-0.10}$ \\
\noalign{\smallskip}\hline\hline\noalign{\smallskip}
$\Sigma(\mbox{H}_2)$ & $20.37^{+0.01}_{-0.01}$ & $20.36^{+0.01}_{-0.01}$ & $20.43^{+0.01}_{-0.01}$ & $20.26^{+0.01}_{-0.01}$ & - & $20.44^{+0.01}_{-0.02}$ & - & $20.33^{+0.01}_{-0.01}$ & - & $20.50^{+0.01}_{-0.01}$ \\
$\Sigma(\mbox{HD})$  & $14.72^{+0.09}_{-0.10}$ & $14.65^{+0.06}_{-0.05}$ & $14.69^{+0.03}_{-0.02}$ & $14.59^{+0.05}_{-0.05}$ & - & $14.66^{+0.07}_{-0.06}$ & - & $14.71^{+0.09}_{-0.08}$ & - & $14.80^{+0.08}_{-0.08}$ \\
\noalign{\smallskip}\hline\noalign{\smallskip}
\end{tabular}

\raggedright
$^1$from fit, $^2$from $W_\lambda(\mbox{CO C-X @}1088\mbox{\,\AA})$; 
see Sect.~\ref{Sect_CDs_Molecules}
\end{table*}


\section{Discussion}

\label{sect_res}

The column densities resulting from the fit are listed in Table~\ref{tab_CDs}.
If a species is not detected we give $3\sigma$ upper limits determined at the
positions of strong, but not neccessarily the strongest, unblended transitions.

\subsection{Velocity structure}

We find one prominent absorption component, and in some spectra a second one
bluewards which is weaker but clearly distinguishable.

The first, stronger velocity component is seen in H$_2$ absorption in all 
spectra at heliocentric velocities of about $-8$~km\,s$^{-1}$.
Velocities can be fitted with a precision of $1$~km\,s$^{-1}$ or below.

We observed that the velocity of some metal absorption lines is different from
that of the molecules by up to $6$~km\,s$^{-1}$ bluewards.
This is observable only in those spectra which show a single velocity
component.
In the other spectra it is probably obscured by blendings of the components.
We fitted these lines using slightly different velocity offsets.
In two cases the \ion{O}{i} transition at 1039~\AA\, is seen to be shifted by
another $\sim -4$~km\,s$^{-1}$.

The origin of this shift may be unresolved substructure in the form 
of an additional blueward component that is visible only in metal lines.
A complex velocity substructure is known to exist from ultra-high resolution 
optical spectra of sight lines in this general direction 
\citep{crawford95,crawford01}.
A problem in the FUSE wavelength calibration would not be restricted 
to the transitions of metals and thus can be ruled out as a cause for such a 
shift.

The second, less prominent velocity component is found in at least three of 
the spectra (\object{HD\,152249}, \object{HD\,152270}, \object{HD\,326329}), 
mainly in the weaker metal transitions at a velocity of 
about $-40$~km\,s$^{-1}$.
It should be noted that these three sight lines are located closely together
at the eastern edge of the target field.
On other sight lines a second component may also be present at that velocity 
judged from well-defined metal lines like \ion{P}{ii}\,(1152). 

In optical ultra-high resolution spectra, \citet{crawford01} 
resolves numerous velocity components in \ion{Na}{i} and \ion{K}{i}, 
with stronger column density peaks around 
$-39$~km\,s$^{-1}$, $-9$~km\,s$^{-1}$, and $+2 $~km\,s$^{-1}$. 
\citet{crawford95} finds numerous absorptions of CH, CH$^{+}$, and CN,
again peaking around velocities of $-9$~km\,s$^{-1}$ and $+2 $~km\,s$^{-1}$.
Especially the velocity of the $-9$~km\,s$^{-1}$ component varies, however, 
depending on the environment the absorbing element lives in.

\label{mention_CO}
CO emission line surveys for this general direction show two main velocitiy
components as well \citep[e.g.,][]{bronfman89}.
However, the component velocity separation in that data is slightly larger than
in our spectra (about $40$~km\,s$^{-1}$ compared to our $32$~km\,s$^{-1}$), and
the velocities are shifted by $-10$~km\,s$^{-1}$. 
As supposed by \citet{bronfman89}, the emission at negative velocities 
should originate in the far background of \object{NGC\,6231}.
It is therefore not related to the absorption in the NGC\,6231 star spectra,
despite the similar velocity.
The emission components seen at positive velocities can be associated with the
Lupus molecular cloud region at a distance of about $150$~pc 
\citep{crawford00b}, and those may indeed interfere with our sight lines.

\subsection{Column densities}

\begin{figure*}[t!]
\sidecaption
  \includegraphics[width=12cm]{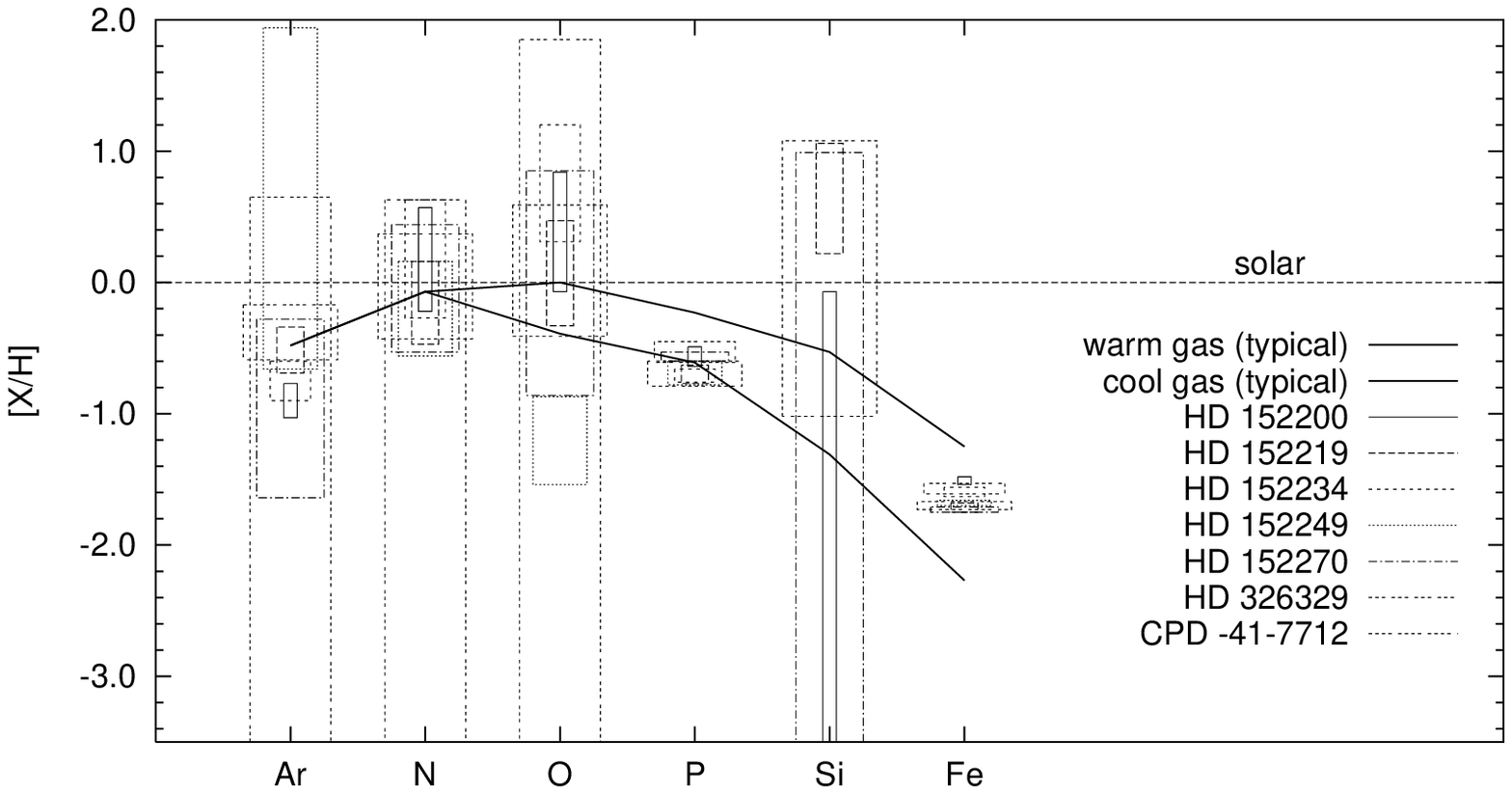}
  \caption[]{\hspace{1mm}Element depletions as derived from the seven sight
             lines observed.
             The width of the error boxes varies for better readability only.
             The bold lines represent the values typically found in the 
             interstellar medium in warm ({\it upper bold line}) and cool  
             ({\it lower bold line}) gas \citep{savage96}.
             The scatter especially in Ar, O, and Si is due to difficulties in 
             determining the column densities for these metals.
            }
  \label{fig_depletion}
\end{figure*}

\subsubsection{Molecules}

\label{Sect_CDs_Molecules}

H$_2$ can be fitted with high accuracy due to the multitude of transitions 
covering a large range of transition probabilities. 
Only the middle excitation states (esp. $J=2,3$) are less accurate, resulting 
from their location on the flat part of the curve of growth (see above, 
Sect.~\ref{sect_obs}).
A mean $N(\rm{H}_2)\simeq20.38$ indicates a sizeable amount, in line
with the values of $E(B-V)$.
The higher excitation levels $J=6$ and $7$ which are highly desired for
excitation modelling are hardly separated from the noise, though.

Also the HD absorption is, in the $J=0$ state, strong enough for a clean fit.
The $J=1$ state is close to the detection limit but may be visible in some 
cases. 
We give upper limits for the HD $J=1$ column density.

Fitting CO is more difficult.
The resolution of FUSE is too low to resolve the individual transitions in the
CO absorption bands.
Typically, the Doppler $b$ value in an environment where CO is found is low.
This means that the equivalent widths are not neccessarily located on the
linear part of the curve of growth; they are thus no longer additive.
The column densities can therefore only be lower limits.

The CO abundances along our sight lines are, with respect to the H$_2$ 
abundances, within the usual range. 
CO column densities in the ISM typically increase significantly at H$_2$
column densities $>10^{19}$~cm$^{-2}$.
The ratio of $N(\rm{CO})/N(\rm{H}_2)$ found fits into the relation drawn by 
\citet{federman80}. 
However, the ratio of $N(\rm{CO})/N(\rm{H}+2\rm{H}_2)$ deviates from other 
Galactic sight lines, presumably due to the fact that the total sight line 
samples large amounts of \ion{H}{i} in regions where no molecular gas, and 
especially no CO, exists.

The molecule column densities, of H$_2$ as well as of HD and CO, 
are not automatically linearly related to those of \ion{H}{i},
because they have their origin in a far more constrained fraction of the sight
line.
Molecular gas is in general confined in the denser regions of the interstellar
clouds and is thus of a more clumpy nature.

\subsubsection{Neutral hydrogen}

The deviation of the column densities of \ion{H}{i} is of prime importance. 
They are the basis for the depletion values and thus any analysis of their 
variation, as well as for any statement on a possible abundance variation
between the different sight lines.
Judged from the relative abundances of the, in general, only marginally 
depleted P and of Fe \citep[see, e.g., the review by][]{deboer87a}, 
all our \ion{H}{i} abundances are consistent with the typical P and Fe values
observed along high column density sight lines (Fig.~\ref{fig_depletion}).
The values towards \object{HD\,152234} and \object{HD\,152249} fully
agree with the results from the IUE survey of \citet{diplas94} as well.

We observe a variation in the total gas column density, 
$N_{\mathrm{tot}}=N(\mathrm{\ion{H}{i}})+2\,N(\mathrm{H}_2)$, between 
$\log N_{\mathrm{tot}}=21.34$ (\object{HD\,152200}) and $21.52$ 
(\object{HD\,152270}), which is about $50\%$ over an angle of $5\farcm7$ 
($60\%$ if one only takes \ion{H}{i} into account).
The variation is rather a linear increase from the western to the eastern
edge of the field of view than a fluctuation due to structure on even smaller 
scales.
The expected large number of \ion{H}{i} absorbers along the sight lines will,
even if they are spatially structured, smear out any fluctuation, leaving just
a general smooth increase of column density.
This increase in $N_{\mathrm{tot}}$ is in agreement with the map from IRAS for
this region, which shows a smooth rise in dust emission in the same general 
direction (Fig.~\ref{fig_positions}).
The effect may also be connected to the second velocity component that we 
observe towards the group of targets located in the north-eastern region of 
our field.

\subsubsection{Metals}

\label{discuss_metals}

The column densities derived for the metals are given in Table~\ref{tab_CDs}.
\ion{Fe}{ii} is the only metal with a large number of usable transitions
in the FUSE spectral range, which results in a reliable fit and the smallest
uncertainties in column density.

Also \ion{P}{ii} seems, with only two transitions 
($\lambda\lambda$ $1124$, $1152$~\AA), 
to give good results if one compares its derived depletion with typical values
\citep{deboer87a}. 
To get proper metal depletions we use the metal column densities summed over
all velocity components, since we are unable to separate any velocity 
components in \ion{H}{i}. 
This, however, reduces the value of the depletions for the interpretation of
the physical conditions in a particular velocity component.

The \ion{C}{i} ground state transitions blend with 
various transitions of the excited states of neutral carbon and are difficult
to unravel.
Therefore, upper limits are listed in most cases.
Information on \ion{C}{ii}, which is the dominant state of carbon in neutral
gas, is not available in the FUSE spectral range.

For some important ions the number of usable transitions in the FUSE spectral
range is small (\ion{N}{i} $\lambda\lambda$ $1134.165$, $1134.415$, $1134.980$,
1159.817; \ion{O}{i} $\lambda$ $1039.230$; \ion{Si}{ii} $\lambda$ $1020.699$), 
and we cannot rule out stellar contamination.
We therefore have to assume large errors for these column densities.

The accuracy of the \ion{Ar}{i} abundance is compromised due to difficulties
in fitting the only two transitions. 
The second velocity component at $-40$~km\,s$^{-1}$ seems to be present on
all sight lines for \ion{Ar}{i}, which has an effect on the $b$ value to
choose for the fit.
Also, the \ion{Ar}{i} lines are located on the flat portion of the curve of 
growth, making a column density determination highly sensitive to the chosen 
$b$ value.

\ion{O}{i} is especially hard to fit on the two-component sight lines, 
in particular towards \object{HD\,152249} and \object{HD\,152270}.
Fits are possible only to one flank of the profile, which requires a precise
wavelength calibration at the position of the transition.
Table~\ref{tab_CDs} states the results from the fit for \ion{O}{i}, 
but the errors given are only the results from the fitting process itself and
cannot represent the larger true uncertainty.


\section{Interpretation}

\label{sect_interp}

\subsection{Sight line structure}

\object{NGC\,6231} is located in the centre of the \object{Sco OB1} 
association, at the near side of the Sagittarius spiral arm.
The structure of the interstellar medium along the line of sight is complex.

The extinction survey of \citet[][field 343/+1]{neckel80} shows in this 
direction a sharp increase in interstellar extinction by $A_V=0.75$~mag that 
takes place between about $100$ and $300$~pc distance.
As mentioned before (Sect.~\ref{mention_CO}), the sight line passes the outer
regions of the Lupus molecular cloud region,
an extended star forming region at the near edge of the Sco-Cen OB association.
The \element[][12]{CO} emission maps of \citet{bronfman89} show that our sight
line misses this CO cloud by $1\degr$, only about $2.6$~pc at a distance
of $150$~pc.

Another sharp increase in extinction is found by \citet{neckel80} at a distance
of 700-1000~pc, again by $A_V=0.75$~mag.
Beyond this point, the sight line seems to be clear up the cluster at about 
$1900$~pc.

The heliocentric radial velocity of \object{NGC\,6231} is about 
$-25$~km\,s$^{-1}$, depending on the method used in the determination
\citep[e.g.,][]{laval72, dambis01}.
Gas surrounding the association is expected to have been cleared away, so its 
absorption velocity will be more negative.
In consequence, we can assume that gas at more positive absorption velocities 
is not associated with the cluster but located somewhere on the line of sight.

\begin{table}[t!]
\centering
\caption{H$_2$ Boltzmann temperatures $T_{01}$ and equivalent excitation 
         temperatures $T_{35}$ for the $-8$~km\,s$^{-1}$ component.
        }
\label{tab_h2temp}
\begin{tabular}{lcl}
\hline\hline\noalign{\smallskip}
Target & $T_{01}$ [K] & \multicolumn{1}{c}{$T_{35}$ [K]} \\
\noalign{\smallskip}\hline\noalign{\smallskip}
\object{HD\,152200}     & $66\pm2$ & $309\pm13$ \\
\object{HD\,152219}     & $71\pm3$ & $259\pm11$ \\
\object{HD\,152234}     & $64\pm2$ & $286\pm11$ \\
\object{HD\,152249}     & $70\pm5$ & $235\pm6 $ \\
\object{HD\,152270}     & $65\pm4$ & $267\pm41$ \\
\object{HD\,326329}     & $69\pm3$ & $216\pm1 $ \\
\object{CPD\,-41\,7712} & $68\pm3$ & $204\pm10$ \\
\noalign{\smallskip}\hline
\end{tabular}
\end{table}

Following this reasoning, the velocity component seen at $-40$~km\,s$^{-1}$ 
probably originates in the outwards-moving shell of \object{Sco OB1}. 
Ultra-high resolution optical spectra towards \object{HD\,152249} 
\citep{crawford01} show substructure in this velocity component that is not 
resolved by FUSE.

It seems reasonable to assume that the absorption at $-8$~km\,s$^{-1}$ 
is associated with the outer regions of the Lupus molecular clouds where
strong absorption is expected to happen.
However, velocities at this longitude do not provide, using Galactic rotation
models, any distance information that would back up this assumption. 
The detection of the second velocity component at 
$v_{\rm rad}=-40$~km\,s$^{-1}$ in only a subgroup of the sight lines may also
point to a location of that component in the background, near \object{Sco OB1}.

The slightly velocity-shifted metal lines then may result from an unresolved
blend of absorption at the Lupus cloud distance and at larger distances, maybe
including even more unresolved components not containing H$_2$.
The extinction observed between $700$ and $1000$~pc may be connected to these
unresolved components.

\subsection{Physical parameters of the gas}

\label{sect_physparm}

\begin{table*}[t!]
\centering
\caption{Total column densities and logarithmic column density ratios for the 
         molecules in the $-8$~km\,s$^{-1}$ component.
        }
\label{tab_mol_ratios}
\begin{tabular}{lrrrrrrrr}
\hline\hline\noalign{\smallskip}
 & \object{HD\,152200} & \object{HD\,152219} & \object{HD\,152234} & \object{HD\,152249} & \object{HD\,152270} & \object{HD\,326329} & \object{CPD\,-41\,7712} & \multicolumn{1}{c}{Mean} \\
\noalign{\smallskip}\hline\noalign{\smallskip}
$E(B-V)$ &  0.43 & 0.46 & 0.45 & 0.44 & 0.46 & 0.44 & 0.46 \\
\noalign{\smallskip}\hline\noalign{\smallskip}
$\Sigma(\mbox{H}_2)$ & $20.37^{+0.01}_{-0.01}$ & $20.36^{+0.01}_{-0.01}$ & $20.43^{+0.01}_{-0.01}$ & $20.26^{+0.01}_{-0.01}$ & $20.44^{+0.01}_{-0.02}$ & $20.33^{+0.01}_{-0.01}$ & $20.50^{+0.01}_{-0.01}$ & \\
$\Sigma(\mbox{HD})$  & $14.72^{+0.09}_{-0.10}$ & $14.65^{+0.06}_{-0.05}$ & $14.69^{+0.03}_{-0.02}$ & $14.59^{+0.05}_{-0.05}$ & $14.66^{+0.07}_{-0.06}$ & $14.71^{+0.09}_{-0.08}$ & $14.80^{+0.08}_{-0.08}$ & \\
$\log f$ & $-0.66$ & $-0.79$ & $-0.69$ & $-0.90$ & $-0.78$ & $-0.80$ & $-0.67$ \\
\noalign{\smallskip}\hline\noalign{\smallskip}
H$_2$/HD & $5.65^{+0.09}_{-0.10}$ &  $5.71^{+0.06}_{-0.05}$ &  $5.74^{+0.03}_{-0.02}$ &  $5.67^{+0.05}_{-0.05}$ & $5.78^{+0.07}_{-0.06}$ & $5.62^{+0.09}_{-0.08}$ & $5.70^{+0.08}_{-0.08}$ & $5.70\pm0.05$\\
H$_2$/CO & $6.52^{+0.06}_{-0.12}$ & $6.58^{+0.08}_{-0.04}$ & $6.52^{+0.03}_{-0.02}$ & $6.44^{+0.05}_{-0.01}$ & $6.49^{+0.05}_{-0.03}$ & $6.34^{+0.11}_{-0.05}$ & $6.61^{+0.11}_{-0.06}$ & $6.50\pm0.09$ \\
\noalign{\smallskip}\hline
\end{tabular}
\end{table*}

\begin{figure}[t!]
\resizebox{\hsize}{!}{\includegraphics{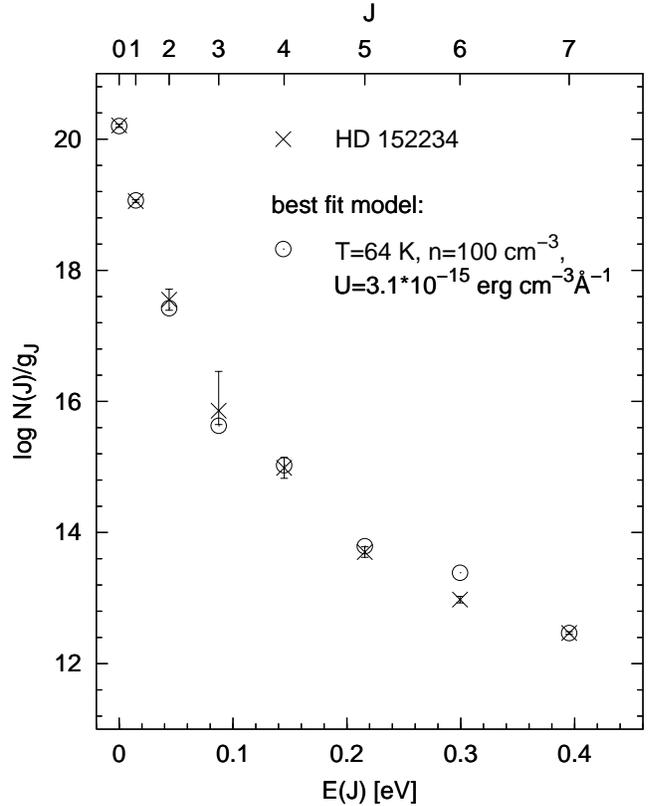}}
\caption[]{Measured H$_2$ excitation and excitation model for 
           \object{HD\,152234. 
           Testing of various models restricts the parameters to a hydrogen
           density $n$ of not more than $1000$~cm$^{-3}$ and a radiation 
           field at $\lambda\simeq1000$~\AA\, with 
           $U_{\lambda}=(3.1\pm0.5)\cdot10^{-15}$\,erg\,cm$^{-3}$\AA$^{-1}$}.
           Shown is the fit for $n=100$~cm$^{-3}$.
          }
\label{fig_excite}
\end{figure}

The excitation of the H$_2$ molecule results from two separate mechanisms:
The lower levels are predominantly excited by collisional excitation
that is described by a Boltzmann temperature. 
The higher levels (from $J=2$ on upwards) are populated mainly by molecules
cascading back from even higher excitation states after either being pumped
to these states by UV photons, or in the molecule formation process.
This excitation is in general described by fitting another Boltzmann 
distribution characterised by a so called equivalent excitation temperature.
The low level H$_2$ excitation temperatures for the individual sight lines 
are given in Table~\ref{tab_h2temp}. 

We did extensive excitation modelling of H$_2$ for some of the sight lines.
\label{ch_model}
Appendix~\ref{app_model} gives a more detailed description of the model.
In short, the model assumes a single isothermal cloud of constant density and 
the equilibrium population of the rotational levels $J=0$ to $J=7$ for a given 
total H$_2$ column density is calculated.

Using simplifications such as a constant interstellar radiation field and an
approximate treatment of the UV attenuation by dust inside the cloud, the 
models allow an estimate of the level of the FUV radiation field and of the
hydrogen density.

For most stars it is not possible to fully fit the excitation using a single 
H$_2$ component model.
Interestingly, this may again point to a hidden two component structure in the 
gas, with physically distinct environments.

As it was shown, e.g., by \citet{flower86}, the shock front of an expanding 
shell can also excite H$_2$ in a way that is comparable to relatively 
high Boltzmann excitation temperatures.
There are, however, indications against a current shock in the 
\object{Sco OB1} shell \citep{crawford01} and also no indications for a shock
in the Lupus clouds \citep{crawford95}.
The \ion{Na}{i}/\ion{Ca}{ii} ratios indicate a previous shock in the 
\object{Sco OB1} environment that released part of the Ca bound in dust grains,
but since then the gas has cooled down again.

Gas in the \object{Sco OB1} shell, however, would have a different, more
negative, velocity and thus would not hide in the main velocity component.
The most plausible explanation for a hidden second component is found in
either substructure inside the Lupus cloud region or in the gas causing the 
strong extinction at $700$ to $1000$~pc distance.
From our data it is not possible to make any definitive statement on this or on
any even more complicated velocity structure.

The excitation on the HD\,152234 sight line allows a fit of a one component 
model (Fig.~\ref{fig_excite}).
Fitting a range of models constrains the gas density $n$ to not more than 
$1000$~cm$^{-3}$ and the radiation field to 
$U_{\lambda}(1000\,$\AA$)=(3.1\pm0.5)\cdot10^{-15}$\,erg\,cm$^{-3}$\AA$^{-1}$.
This value is about $100$ times the typical Galactic disk radiation field 
\citep[e.g.,][]{habing68}, which could hint at a significant source of 
radiation closeby.

Another estimate of the conditions in the gas is provided by the ratio of the 
two lowermost excitation levels of HD.
In contrast to H$_2$, transitions between even and odd rotational excitation
states are possible.
The population of the $J=0$ and $J=1$ states of HD therefore depends not only 
on collisional but also on excitation by the prevailing radiation field.

However, if one assumes the temperature, density, and radiation field derived 
from the H$_2$ excitation to be valid also for HD, one would get a much higher 
population of the HD $J=1$ level than is observed 
($N(\mathrm{HD}_{J=1})$/$N(\mathrm{HD}_{J=0})\ga0.45$ instead of the observed
$0.2$).
The ratio that is observed constrains the radiation field in the gas to not 
more than $U_{\lambda}\simeq1.4\cdot10^{-15}$~erg\,cm$^{-3}$\,\AA$^{-1}$ 
(or $I<35$), which is at least a factor $2$ smaller than the value from the 
H$_2$ models.

To explain this discrepancy, one has to recall that the H$_2$ excitation model
assumes a single cloud.
A high radiation field is needed to penetrate this large cloud and produce the 
observed excitation pattern.
A number of smaller clouds with the same overall column density along the
sight line would show an excitation pattern similar to the observed one already
at a smaller radiation field, provided the radiation is not attenuated by other
cloudlets.

We tested additional models for a series of clouds with only a fraction of the
total column density each.
The observed excitation pattern is equally well fitted by two or four smaller
clouds and by a radiation field smaller by the same factor, plus a smaller
$b$ value for each cloud.
Such a model would still be consistent with the larger number of absorption 
lines and the correspondingly low $b$ values \citet{crawford01} finds from 
optical ultra-high resolution spectra.

This suggests a considerably smaller radiation field than required by the 
simple one-cloud model (possibly not much higher than the Galactic radiation
field) and a number of smaller distinct clouds along the sight line.

The modelling of the H$_2$ excitation furthermore allows us to estimate the 
formation rate of H$_2$ in the gas. 
The H$_2$ dissocitation rate from the one-component model is 
$1.2\cdot10^{-13}$~s$^{-1}$.
This would require a formation rate coefficient 
$R\simeq1.6\cdot10^{-16}$~cm$^3$\,s$^{-1}$ (for a density of $n=100$~cm$^{-2}$)
to maintain equilibrium between formation and dissociation of the H$_2$.
However, the commonly accepted value for the H$_2$ formation rate is 
$1$-$3\cdot10^{-17}$~cm$^3$\,s$^{-1}$, depending mainly on the composition of
the interstellar dust.
Provided that the H$_2$ formation takes place at only the usual rate, the 
local radiation field will then dissociate the H$_2$ gas within a time scale of
roughly $10^6$~a.
A stable condition might exist if the gas density is as high as 
$\sim500$~cm$^{-3}$, thus lowering the required formation rate coefficient.

\cite{crawford89} gives an analysis of the CH column densities towards 
\object{Sco OB1}.
Using a simple model with an assumed density of $n\ge100$~cm$^{-3}$, he
reproduces these column densities, but the model fails to reproduce the 
CH$^+$ abundance.
Provided that H$_2$ and CH coexist in the same region of space, this 
constrains the density in the molecular gas to be in the range of
$100$~cm$^{-3}\la n\la 1000$~cm$^{-3}$.

The quality of most of the metal abundances is too low for any constraints
on physical gas parameters.

\subsection{Molecular abundances}

The ratios of the molecular species, H$_2$/HD and H$_2$/CO, are, within the 
range of the errors (Table~\ref{tab_mol_ratios}), 
constant over the sight lines.
This allows a good estimate of these astronomically important parameters.

We get a mean $\log N(\rm{H}_2)$/$\log N(\rm{CO})$ of $6.50\pm0.09$.
Although the physical conditions are different, this value can be compared
with results from other models, e.g. from 
\citet[][especially their model H1]{vandishoek88} for the 
\object{$\zeta$~Oph} sightline. 
As expected, this locates the \object{NGC\,6231} gas at the low density edge 
of these translucent cloud models, at a radiation level probably near
the Galactic disk average \citep[e.g.,][]{draine78} and densities of about
$500$~cm$^{-3}$.

Also, we obtain a mean $\log N(\rm{H}_2)$/$\log N(\rm{HD})$ of $5.70\pm0.05$. 
This value is not related to the H/D ratio in a linear way but it also depends,
due to the formation processes involved, on additional parameters like the
proton density in the gas.
Also, in all dominant processes involved in the production of HD ionised 
hydrogen or deuterium plays a role.
For this reason the ratio of H$_2$/HD leads to an estimate for the ionisation
level in the molecular cloud.
Assuming a density of $n=100$~cm$^{-3}$ and an optical depth at $1000$~\AA\, of
$\tau_{\mathrm{UV}}\simeq2$ we get a fraction of ionised hydrogen of 
$x(\rm{H}^+)\simeq3\cdot10^{-5}$.
We assume $N($D$)/N($H$)=1.2\cdot10^{-5}$, the mean value from recent
measurements \citep{moos02,hoopes03,bluhm99} on sight lines beyond the Local 
Bubble.
This low ion fraction, however, does not allow conclusions on the abundance
of free electrons in the gas, which are mainly provided by \ion{C}{ii} under 
the prevailing conditions.

\subsection{Excited neutral carbon}

The excitation of neutral carbon and its dependence on gas density and 
temperature is described by \citet{deboer74b}.
The ratio of \ion{C}{i}\,$(^3P_1)$/\ion{C}{i}\,$(^3P_0)$ is probably
$<-0.2$~dex for our gas, on the \object{HD\,152234} sight line roughly
$-0.2$~dex.
With this value and the temperature of $64$~K that was derived above 
(Sect.~\ref{sect_physparm}) for the H$_2$ gas, one gets a gas density of 
$n_{\rm{H}}\simeq130$~cm$^{-3}$.
Alternatively, a gas density of $100$~cm$^{-3}$ as derived from the H$_2$
models would require a gas temperature of $T\simeq76$~K for the observed
\ion{C}{i} excitation.
In general, the gas pressure $n_{\rm{H}}T$ from \ion{C}{i} is around
$8000$~K\,cm$^{-3}$.

\subsection{Depletions}

The depletion of the metals seems, with the uncertainties in mind, not
unusual. 
The values for phosphor and iron are in the range that is typical for cool disk
gas.
For \ion{Ar}{i} one might, from the depletion diagram 
(Fig.~\ref{fig_depletion}), crudely estimate an additional underabundance by 
$-0.2$~dex due to partial ionisation.
This would correspond with an ionisation rate of the gas of $0.07$, which 
would require a photoionisation rate of $\Gamma($H$)=2.2\cdot10^{-16}$~s$^{-1}$
(assuming $T=5000$~K and $n_{\rm{H}}=0.1$~cm$^{-3}$) or,
density independent, $\Gamma($H$)/n_{\rm{e}}=3.1\cdot10^{-14}$~s$^{-1}$.
However, the uncertainty in the \ion{Ar}{i} column density is large 
(as discussed in Sect.~\ref{discuss_metals}) and these rates should not be 
overinterpreted.

\subsection{Effects of velocity substructure}

In all these considerations we have to bear in mind that there is velocity 
substructure in our absorption lines which we are unable to resolve,
and that there is probably also absorption at similar velocities from
physically distinct regions in the ISM.
The gas seen in the partially ionised \ion{Ar}{i} is surely distinct from the
gas seen in the molecules, the \ion{Ar}{i} being in the outer boundaries of 
denser regions or possibly warmer intercloud medium.
Also the number of molecular clouds is unclear, or its separation into smaller
cloudlets along the line of sight. 
The CO is expected to exist in the shielded cores of these clouds, so its
density will be somewhat higher than that of the H$_2$ gas.
Nevertheless, all methods and all models seem to point at an extended 
substructure of small cloudlets which can be deeply penetrated by the 
surrounding radiation field.

\subsection{Spatial substructure}

Besides the  structure of the gas along the lines of sight, a main aspect of 
this paper is the structure over the field of view, perpendicular to the sight
lines. 
Of the metals only \ion{Fe}{ii} and \ion{P}{ii} provide useful information
on this structure.
To get a valid comparison with the \ion{H}{i} abundance that is not resolved in
velocity, one has to use the column density sums over the velocity components
observed, which results, again, in a lack of significant variation for both
species.
Thus, the statistical blend of different absorbers hides any information on
column density fluctuations over a particular cloud.

The molecular gas may, however, provide some insight since it very likely is
more sparsely distributed along the sight lines.
We observe a rather irregular variation of the H$_2$ column density over
the field of view, with variations of $40$\% over a scale of roughly $1\arcmin$
(\object{HD\,152219} and \object{CPD\,-41\,7712} being the extremes).

The angular separation of these two sight lines corresponds to only 
$\sim0.05$~pc at the distance of $150$~pc assumed for the molecular gas.
Under the -- simplifying -- assumption of circular H$_2$ cloudlets, one 
derives a radius of not more than $0.1$~pc ($2\farcm3$) for the strongly 
absorbing H$_2$ cloud in front of \object{CPD\,-41\,7712}, at the edge of the
triangle formed by \object{HD\,152200}, \object{HD\,152219}, and 
\object{HD\,152249}.

The difference in column density in this region is about 
$9\cdot10^{19}$~cm$^{-2}$. 
With a radius of $0.1$~pc such a cloutlet would then have an average density
of roughly $300$~cm$^{-3}$, again in accord with the values derived above.

Some variation is also visible in the equivalent excitation temperatures that 
are connected to the intensity of the radiation field at the location of the 
gas.
These temperatures, however, are only slightly anticorrelated with the 
H$_2$ column density.

The imperfections in the calibration of the FUSE spectra do not allow us to
make any hard statement on potential velocity substructure over the field of
view. 
From the agreement of the individual H$_2$ absorption velocities we assume 
a quite homogeneous velocity distribution of at least the H$_2$ gas over our 
field of view. 

Absorbers seen in metal lines are far more abundant along the sight lines, so
the absorption is seen at many more different velocities. 
The variation seen in the velocities of these lines probably rather comes from 
faint variations in the blends of different absorbers than from real velocity
substructure.


\section{Concluding remarks}

\label{sect_conc}

We find a main absorption component at a velocity of $-8$~km\,s$^{-1}$, 
and a weak second one at about $-40$~km\,s$^{-1}$ that is resolved only on some
of the sight lines.
The main component is probably associated with the Lupus molecular cloud region
at $150$~pc distance, while the second one may show gas in the Sco OB1 
shell.

For the molecular gas we find strong arguments for a density of the order of 
magnitude of $100$~cm$^{-3}$.
The radiation field $U_{\lambda}$ in the molecular cloud region may be strong,
if one considers single-cloud H$_2$ excitation models. 
The CO abundance rather hints to a radiation field near the mean Galactic
value, which could also be consistent with multi-cloud models for the H$_2$ 
excitation.
At a high level of radiation in the clouds the molecular gas may be in 
a process of dissolution.

In addition to this indicated clumpiness along the sight lines
the gas structure perpendicular to the sight lines is analysed.
The spatial variation in the column densities for the Lupus cloud gas is
found to be $40\%$ for H$_2$ and $60\%$ for \ion{H}{i}. 
This strongly indicates clumpiness in the gas on scales of $0.1$~pc and below.
We are not able to resolve any spatial variation in the velocity of the 
absorbing gas over the field of view.

Clumpiness of the molecular interstellar medium has been detected in 
several recent works and on various scales, e.g., \citet{pan01} on 
sub-parsec scales and \citet{lauroesch00} on scales down to almost ten AU.

This work is another confirmation of the inhomogeneous structure of the 
Galactic ISM on sub-parsec scales.

\appendix
\section{A simple model of H$_2$ excitation}

\label{app_model}

This Appendix gives a concise overview of the H$_2$ excitation model used in
our analysis (Sect.~\ref{ch_model}). 
The full details are found in \citet{bluhm03a}.

The basic properties of the model are:
\begin{itemize}
\item{6 parameters ($N({\rm{H}}_2)$, $b$, $T$, $n_{\mathrm{p}}$, 
$n_{\mathrm{H}}$, $U_{\lambda}$) are used to calculate 8 column densities 
$N(J=0..7)$.
}
\item{Assumptions or approximations are: $U_{\lambda}(\lambda)=const.$,
$T_{\rm kin}(l)=const.$, $n_{\mathrm{H}}(l)=const.$, and 
$n_{\mathrm{p}}(l)=const.$ (where $l$ is the depth along the sight line in the 
cloud).
}
\item{Collisional excitation by particles other than neutral and ionised 
hydrogen is neglected.
}
\item{Collisional excitation and de-excitation of vibrationally excited levels
as well as UV absorption out of vibrationally excited levels are neglected.
}
\item{The radiation which is responsible for the UV-pumping is assumed to come
from the direction of the background source, i.e. the radiation propagates
along the sight line.
}
\item{While self-shielding in absorption lines is included (by integration over
the Voigt absorption profiles), blending of lines from different species
(different $J$-levels of H$_2$ and/or atomic species) is neglected. 
Thus for high column density sight lines, the radiation field will be somewhat 
underestimated.
}
\item{Absorption of the UV continuum within the cloud is taken into 
consideration by applying approximate correction factors to the cascade entry
rates.  
The value of $U_{\lambda}$ refers to the irradiated edge of the cloud.
}
\item{The UV-pumping process is calculated in a simplified manner by using 
averaged ``cascade efficiency factors'' 
\citep[based on the cascade efficiency factors $a(v_0 J_0;J)$ published in][]{black76}. 
The averaged factors $a(J';J)$ give the probability that a molecule in level 
$J'$ before the absorption of an UV photon ends up in level $J$ after the 
UV-pumping process.
}
\item{It is assumed that 87\% of the absorptions lead to UV-pumping while the 
remaining 13\% lead to dissociation \citep[see][]{abgrall92}.
In equilibrium, the total number of newly formed molecules is the same as the
number of destroyed molecules, but the ortho-to-para ratio will differ in
general.
According to theoretical calculations, an ortho-to-para ratio of 3 is probable
\citep[see, e.g.,][]{takahashi01a}.
Since it is unlikely that a newly formed molecule has lost its formation
energy completely when it leaves the dust surface, it will enter the
rovibrational cascade, albeit with a different cascade entry distribution than
the UV-excited molecules.
\citet{takahashi01b} modelled the vibrational and rotational excitation of 
H$_2$ newly formed on icy, silicate, and carbonaceous dust surfaces.
The most populated vibrational levels are $v=7$, $6$, and $2$, respectively, 
with an average rotational excitation of about $J=10$.
As cascade efficiency factors $a_f(J)$ for newly formed molecules mean values
of $a(v_0,J_0;J)$ (for $v_0=2, 6, 7$ and $J_0=9, 10$) have been chosen.
In an analysis of experimental results about the formation of molecular 
hydrogen \citet{katz99} find a non-vanishing probability that a newly formed 
molecule is not immediately ejected from the dust surface.
A molecule lingering on the surface will loose a major fraction of its
formation energy or even be ``thermalized'' before it is eventually desorbed.
To take this into account, for $40$~\% of the newly formed molecules a 
Boltzmann distribution at the gas temperature is assumed.
}
\end{itemize}  
Probably the most uncertain point of the model is the treatment of molecule
formation.                                                             
The rotational levels predominantly affected by this are $J=6$ and $J=7$.


\begin{acknowledgements}

 OM was in part supported by Deutsche Forschungsgemeinschaft (DFG) grant 
Bo~779/24, 
HB in part by the DFG Graduiertenkolleg 
{\sl ``The Magellanic Clouds and other dwarf galaxies''} (GRK 118).
We thank the referee, S.R.~Federman, for helpful suggestions.

\end{acknowledgements}



\bibliographystyle{aa}
\bibliography{h4600}

\end{document}